\documentstyle[prl,aps,twocolumn,eqsecnum,epsf]{revtex}

\draft

\def\Spicture#1#2#3{\begin{tabular}{c}{\epsfxsize=#1\epsfbox{#2}}\\
\mbox{#3}\end{tabular}}

\begin{document}

\twocolumn[
\hsize\textwidth\columnwidth\hsize\csname @twocolumnfalse\endcsname

\title{
\vspace{-3mm}
\hfill {\small Preprint DFPD 00/TH/21, TRI-PP-00-09}
\vspace{2mm}\\
 Why is the three-nucleon force so odd? }

\author{  L. Canton$^1$ and W. Schadow$^2$ }
\address{$^1$Istituto Nazionale di Fisica Nucleare, Sez. di Padova,
Via F. Marzolo 8, Padova I-35131, Italy \\
      $^2$TRIUMF, 4004 Wesbrook Mall, Vancouver, British Columbia,
      Canada V6T 2A3}

\date{June 12, 2000}

\maketitle

\begin{abstract}
By considering a class of diagrams which has been 
overlooked also in the most
recent literature on three-body forces, we extract a new contribution to the 
three-nucleon interaction which specifically acts on the triplet 
odd states of the
two nucleon subsystem. In the static approximation, 
this 3$N$-force contribution is fixed by the underlying 
2$N$ interaction, so in principle there are no free parameters to 
adjust. The 2$N$ amplitude however enters in the 3$NF$ diagram
in a form which cannot be directly accessed or constrained 
by $NN$ phase-shift analysis.
We conclude that this new 3$N$-force contribution
provides a mechanism which implies that 
the presence of the third nucleon modifies
the $p$-wave 
(and possibly the $f$-wave)
components of the 2$N$ subsystem  
in the triplet-isotriplet channels.
\end{abstract}

\vspace{4mm}

]

\section{  Introduction  }
\label{  intro  }

\vspace{-2pt}
The three-nucleon system represents an ideal
testing ground for the nucleon-nucleon interaction\cite{Gloeckle96}.
This testing represented a challenge which required a great deal of
efforts amongst various research groups active in this area in order
to provide a comparison between extremely reliable calculations
and precise measurements. In this field there have been four main
areas of research which proved to be crucial for the progress in our 
understanding of the three-nucleon problem.
These four areas are: 
1) $NN$ Phase-shift analysis. 
2) 2$N$ potentials. 
3) 3$N$ calculations.
4) Experiments on few-nucleon systems.
Recent advancements on these topics can be found in Ref.~\cite{pan3N}.

The combined result of all these studies and efforts has revealed 
that the presence of the third nucleon modifies the interaction
between the remaining two.
A serious difficulty must be faced at this point: namely,
how can we conveniently describe the modifications of the $NN$ force
operated by the third nucleon?
Most likely, we cannot avoid ambiguities in the description of
the interaction modifications by the third nucleon, since
the two-nucleon amplitudes enter off-the-energy-shell
in the three-body quantum mechanical equations.
And it is known \cite{Polyzou90}, that it is possible to introduce
(maybe, unrealistic) modifications in the $NN$ off-shell structure
that mimics the interaction effects due to the presence 
of the third nucleon.
The lesson we must learn from this is that it is not sufficient
to generate a phenomenological $NN$ potential that perfectly
reproduces the most up to date phase-shift analysis.
One must also generate a $NN$ potential by using theoretical insight
as much as possible in order to constrain the off-shell properties
of the $NN$ amplitude. If this is not the case, a $NN$ potential
which fits precisely the experimental data but with the erroneous
off-shell behavior would not provide reliable results for the
3$N$ system, nor can be used to test for the presence of 3$N$ forces.

On the other hand any realistic 2$N$ potential underbinds 
the triton and this is
the first signal that three-body nuclear forces do play a role, 
and that the off-shell ambiguities are not so dramatic
(if the 2$N$ potential is theoretically constrained in its 
one-pion-exchange term and one assumes the requirements of
minimal momentum dependence and/or non-locality). 
The off-shell ambiguities can be seen by the fact
that one must adjust separately with each 2$N$ potential
the parameters of the 3$N$ force in order to reproduce exactly 
the experimental binding energy of the 
triton~\cite{Huber98},
and this suggests that
2$N$ forces and 3$N$ forces must be generated consistently,
within the same theoretical scheme.

Aside for the problem with the triton underbinding,
other signals for possible evidences of 3$NF$ effects
must be sought in the 3$N$ continuum.
One signal arrived recently from the study of the unpolarized
differential cross section for $nd$ elastic scattering
between 60 and 200 MeV.
Continuum calculations with 2$NF$ underestimate
the minimum by a 30\% effect (the Sagara discrepancy).
Rigorous Faddeev calculations with the inclusion of 3$NF$
completely solve the discrepancy between 60 and 140 
MeV~\cite{Witala98}. A coupled-channel calculation with the 
explicit treatment of the $\Delta$-isobar excitation which 
leads to effective 3$NF$`s in the three-nucleon subsystem
reduces the Sagara discrepancy by a considerably 
large fraction\cite{Nemoto98}. Useful insights about the
importance of possible 3$NF$-diagrams should also be
obtained from the study of pion production/absorption
mechanisms in the 3$N$ system, since progress
has been recently made in the theoretical treatment
of these reactions at the $\Delta$ resonance~\cite{Canton97,Canton98b}
and in the threshold region~\cite{Canton99}.

At the present stage, however, the existing 3$NF$ models 
achieved only a limited
success in explaining the discrepancies between theory and experiments,
and it is possible that the spin-isospin
structure of the full three-nucleon force is not well understood, yet.
Indeed, as is well known\cite{Gloeckle96}, the  comparison
between theory and experiments reveals that existing 3$N$ forces
do not provide the correct structure of the vector analyzing 
powers, both for the proton, $A_{y0}$, and for the deuteron
case $iT_{11}$, while the deuteron tensor polarization
observables are described reasonably well. 
This has been evidenced also for $pd$
scattering below breakup threshold\cite{Kievsky96},
where variational techniques based on the
Pair Correlated Hyperspherical Harmonic method 
allowed to incorporate the effects of the Coulomb interaction.
The situation is still the very much same 
in these days as 
has been pointed out also in the most recent Conference on Few-Body 
problems, held in Taipei in March 2000 (FBXVI). 

This puzzling situation about the vector analyzing powers has been
carefully analyzed in two recent publications~\cite{Tornow98,Friar98}.
In Ref.~\cite{Tornow98} it was concluded that, unless 3$NF$ of new structure
could be envisaged, the $NN$ interaction in the $^3P_J$ states
has to be modified in an energy dependent way (i.e. only for 
energies lower than 20 MeV). The study implicitly assumes that
in this energy region 
the results from modern phase-shift analysis\cite{NI93} could be 
possibly corrected in the triplet $p$ waves without affecting 
appreciably the $NN$ data.
In Ref.~\cite{Friar98} all the possibilities for solving
the problem at the level of the two-body interaction 
(especially introducing modifications in the
$^3P_J$ channels) have been attently investigated and then 
ruled out,
with the conclusion that the only viable solution to the
puzzle of the vector analyzing powers must come from a new 
3$NF$ contribution which has not yet been taken into account. 
The authors of Ref.~\cite{Friar98} suggest a 3$NF$ 
of the spin-orbit type, as a possible candidate.

In this paper we discuss the dynamical mechanism
which generate a new component to the three-body force.
This mechanism has been obtained starting from
a formalism~\cite{Canton98} developed
for the treatment of the pion dynamics in the 3$N$ system.
By projecting out the pion degrees of freedom from
that formalism one obtains a 3$N$ dynamical equation
of Alt-Grassberger-Sandhas (AGS) type \cite{AGS}, which incorporates 
the spin-off of the 
pion dynamics beyond that already considered in the 
2$N$ interaction. It has been shown~\cite{Canton2000a} that the 
new pionic terms in the 3$N$ equation can be interpreted as irreducible 
diagrams, contributing to the construction of a 3$NF$.
Herein, we are mainly interested in one specific 3$NF$ mechanism,
which implies an intermediate 2$N$-cluster formation while
a pion is ``in flight". Since 2$\pi$ mesonic retardation effects
do play a role here, we analyzed this role, and found that
they merely act as counter terms, to be subtracted because of 
the presence of a known cancellation effect~\cite{Yang86,Coon86}.
This cancellation effect is correctly taken into account
in the construction of all modern 3$NF$'s~\cite{Weinberg92}.
The novelty of the approach presented herein is that we sized the effect 
of this cancellation more precisely, by allowing the mechanism 
to adjust to the effects of the nuclear medium. This was possible
only because we started from an explicit treatment of the pion 
degree of freedom, while it would have not been possible to see the 
effect within the more common ``instantaneous" approaches to the 
3$N$ force. The nature of these dispersive effects is well known
in approaches devoted to the explicit treatment of the $\Delta$
degrees of freedom\cite{Sauer92}, however a
discussion of the same effects in the presence of an explicit treatment
of the pion dynamics can hardly be found in the literature.
 
From this new 3$NF$ contribution we have extracted in a very natural 
way a specific component
which acts only in the (triplet) odd waves of the 2$N$ subsystem. 
We provide also the first derivation
of the partial-wave expansion of this piece of 
3$N$ force, which can be used in current 3$N$ calculations.
The spin-isospin structure of this 3$N$ force implies that at low energies 
the presence of the third nucleon 
modifies the 2$N$ subsystem in the $^3P_J$
channels, and we suggest that this might
be another possible candidate for 
explaining the inconsistencies registered
between theory and experiments at low energies, at least in those cases
most sensitive to the triplet $p$-waves, such as $A_{y0}$ and $iT_{11}$.

\vspace{-2pt}

\section{ Irreducible 3$N$-force diagrams }

\vfill

The diagrammatic analysis discussed in this section is based
on the systematic method developed in Ref.~\cite{Canton98}
to take into account the pion dynamics into the 3$N$ system.
The method has been originally designed
for the treatment of the 3$N$ dynamics above the pion threshold
and represents an appropriate, connected-kernel, generalization of the
standard Faddeev-AGS three-body equations\cite{AGS,F} for the explicit
inclusion of one meson degree of freedom. In the $\pi$-3$N$
space, the equations are labelled in terms of modified
Yakubovsk{\u\i}-type chains of partitions. 
In the 3$N$ (no-pion) sector, the labelling structure leads
to Faddeev-type components. Specific rules are provided
on how the pion-nucleon vertex interaction couples the
two sectors. By recursive application
of the quasiparticle/separable-expansion method, a modified 3$N$
AGS-quasiparticle equation
is obtained where the explicit pion dynamics is built-in.
This approach has been subsequently considered in 
Ref.~\cite{Canton2000a} where an approximated, practical scheme
for the solution of these equations has been designed.
To the lowest order, the approximation scheme consists simply
in replacing the inelastic components of the $NN$ 
subamplitudes with their leading terms, represented by suitable
combinations of the pion-nucleon vertex interaction.
By setting to zero also these leading terms for
the pion inelasticities in the $NN$ subamplitudes, the $NN$ subamplitudes
become totally elastic and in this limit~\cite{Canton98} 
one re-obtains the standard 3$N$ Faddeev-AGS equation with 2$N$ interactions,
plus a completely disjoint standard Yakubovsk{\u\i}-GS equation\cite{Y,GS} 
for the $\pi$-3$N$ sector. In this limit, the input for the two separated 
three- and four-body equations are the fully elastic $NN$ and $\pi N$ 
$t$-matrices.

In Ref.~\cite{Canton2000a} it is discussed how to project out the pion
degree of freedom from the treatment of Ref.~\cite{Canton98}.  The
result of this procedure, i.e., cooling down the pion from the theory,
can be recast in a very appealing way if one uses a finite-rank
expansion of the elastic $NN$ $t$-matrix. There are methods to
generate these expansions, such as the Ernst-Shakin-Thaler
method~\cite{EST}, and with these methods very reliable and accurate
separable expansions have been generated and
tested~\cite{Cornelius90,Parke91,Schadow98}.

The new 3$N$ equation incorporating the pion dynamics
has the standard AGS form~\cite{AGS}
\begin{equation}
X_{ab}=Z_{ab}+\sum_{c}Z_{ac}\tau_{c}X_{cb} \, ,
\label{AGS-pion}
\end{equation}
where $a$, $b$, and $c$ run over the three Faddeev components
of the 3$N$ system,
and the only modification refers to the driving term,
which can be separated into the following structure
\begin{equation}
Z_{ab}=Z_{ab}^{AGS}+Z_{ab}^{3N} \,.
\label{Z-pion}
\end{equation}
The first contribution is, literally, the standard AGS driving term,
which many groups have been calculating for years, while the second 
term represents
the spin-off of the pion dynamics beyond that already contained in
the 2$N$ interaction. 
Following Ref.~\cite{Canton2000a}, it is possible to analyze
all the pion-exchange diagrams contained in $Z_{ab}^{3N}$
and it is found that they all correspond to irreducible
3$N$ diagrams, and this establishes a link between this $\pi$-3$N$ approach
and those formalisms considering irreducible 3$N$ diagrams
as generators of 3$N$ interactions.

The diagrams that emerged from the analysis of $Z_{ab}^{3N}$
can be classified according to their topological structures,
and we can easily recognize irreducible diagrams that are
well known. But we obtain also other diagrams which
have been overlooked, to the best of our knowledge.

\begin{figure}
\vspace{-0.3cm}
\centerline{\hspace{0.3cm}
\Spicture{2.2 in}{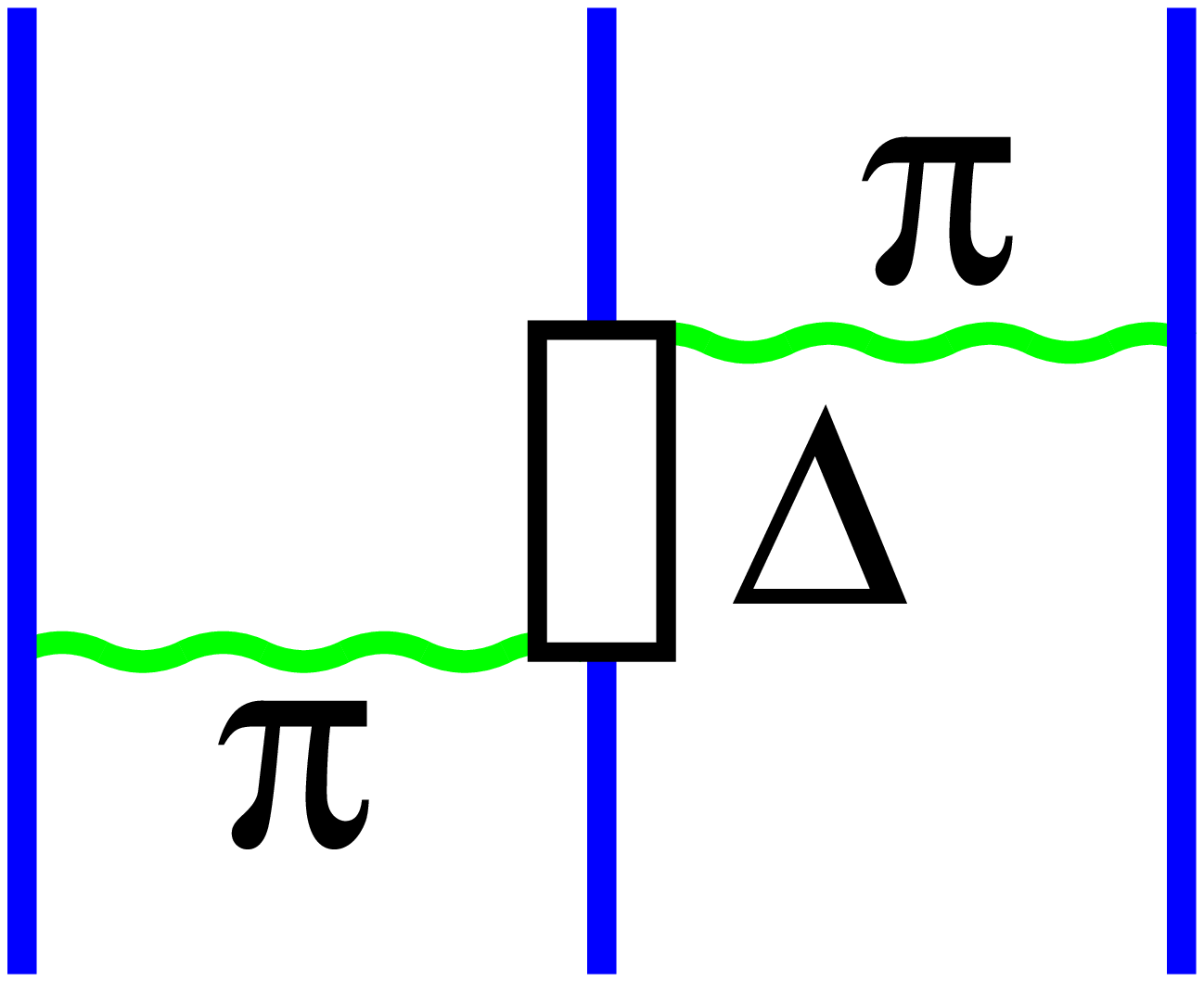}{}
}
\vspace{-1.2cm}
\centerline{\hspace{0.5cm}
\Spicture{2.2 in}{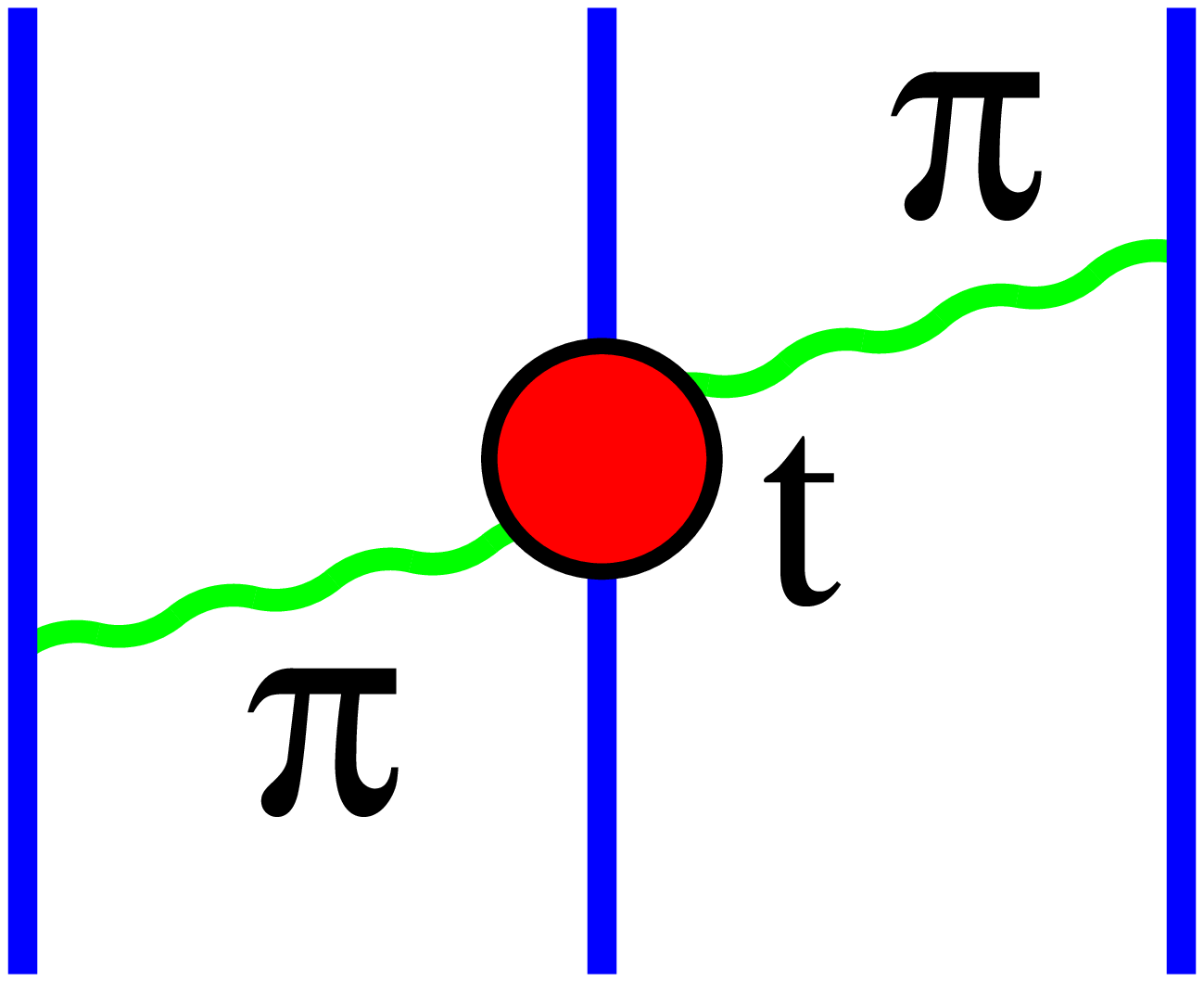}{}
}
\vspace{-0.5cm}
\caption[ ]{Examples of well-known irreducible 3$NF$ diagrams. 
The Fujita-Miyazawa diagram (top) and the non-polar $\pi N$
rescattering diagram (bottom).}
\label{fig1}
\end{figure}

\begin{figure}
\centerline{\hspace{0.3cm}
\Spicture{2.2 in}{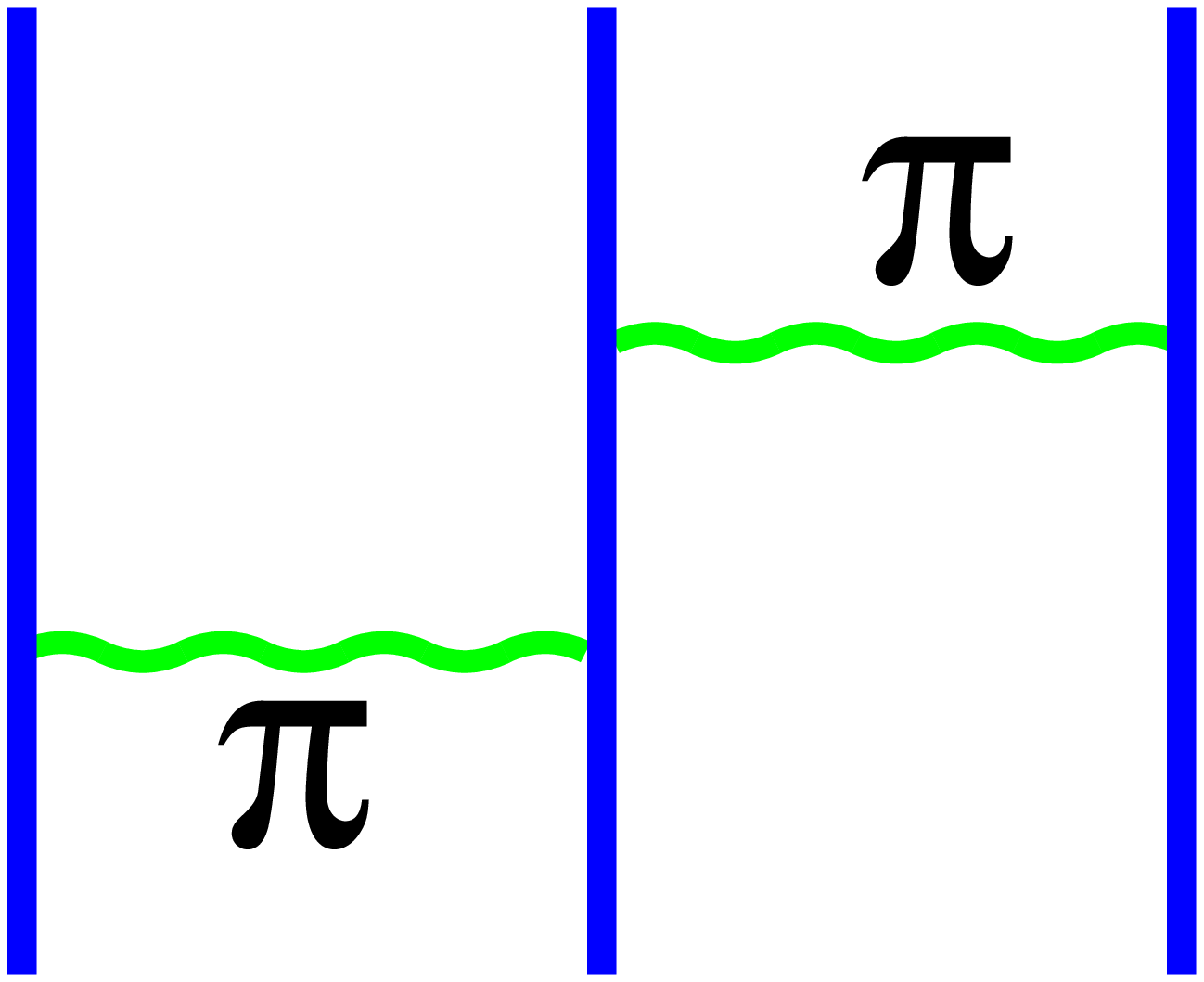}{}
}
\vspace{-0.4cm}
\caption[ ]{The reducible Born diagram. This must be subtracted}
\label{fig1born}
\end{figure}

\vspace{4mm}

The first class of diagram is represented by the diagrams 
reported in Fig.~\ref{fig1},
which describes a pion rescattering by a third nucleon while
being exchanged between the other two. 
As a matter of fact, such a 2$\pi$-exchange structure of the 
3$N$ potential has been
the only one considered in all modern calculations.
This has been pointed out also very recently by Friar {\em et al.}
in Refs.\cite{Huber99,Friar99}. On
the upper part of the figure it is shown
the Fujita-Miyazawa $\Delta$-mediated interaction
\cite{FUMI}, which represents the 
prototype of this topological structure, and accounts for an
important fraction of the generic pion rescattering process. 
In the lower part of Fig.~\ref{fig1} the more general 
pion-rescattering process is exhibited. The ``blob" represents
the $\pi N$ amplitude, where one must subtract
its polar part (corresponding to a nucleon 
propagating in the forward direction) 
to avoid double counting with the nucleonic multiple-scattering 
contributions, since these last are summed up to all orders
in the dynamical 3$N$ equations. The term which 
must be subtracted is  shown in Fig.~\ref{fig1born}
and is called sometimes the (reducible) Born term. The existing
3$N$ potentials
differ mainly in the model calculation of the $\pi N$ $t$-matrix,
whether it is constrained by current algebra and PCAC~\cite{TM},
or inspired by an effective meson-baryon Lagrangian 
constrained by chiral symmetry and current algebra~\cite{BR}, 
or determined by effective field-theoretic methods
involving light-meson dynamics~\cite{Eden96},
or by the more systematic method of $\chi$PT~\cite{TX},
or finally by the form envisaged by the Fujita-Miyazawa 
term~\cite{UA}. Since such contributions (and their short-range 
corrections) have been the subject of very extensive studies, we really 
have nothing to say in addition and will skip to the next classes.
 
\begin{figure}
\centerline{\hspace{1.0cm}
\Spicture{1.9in}{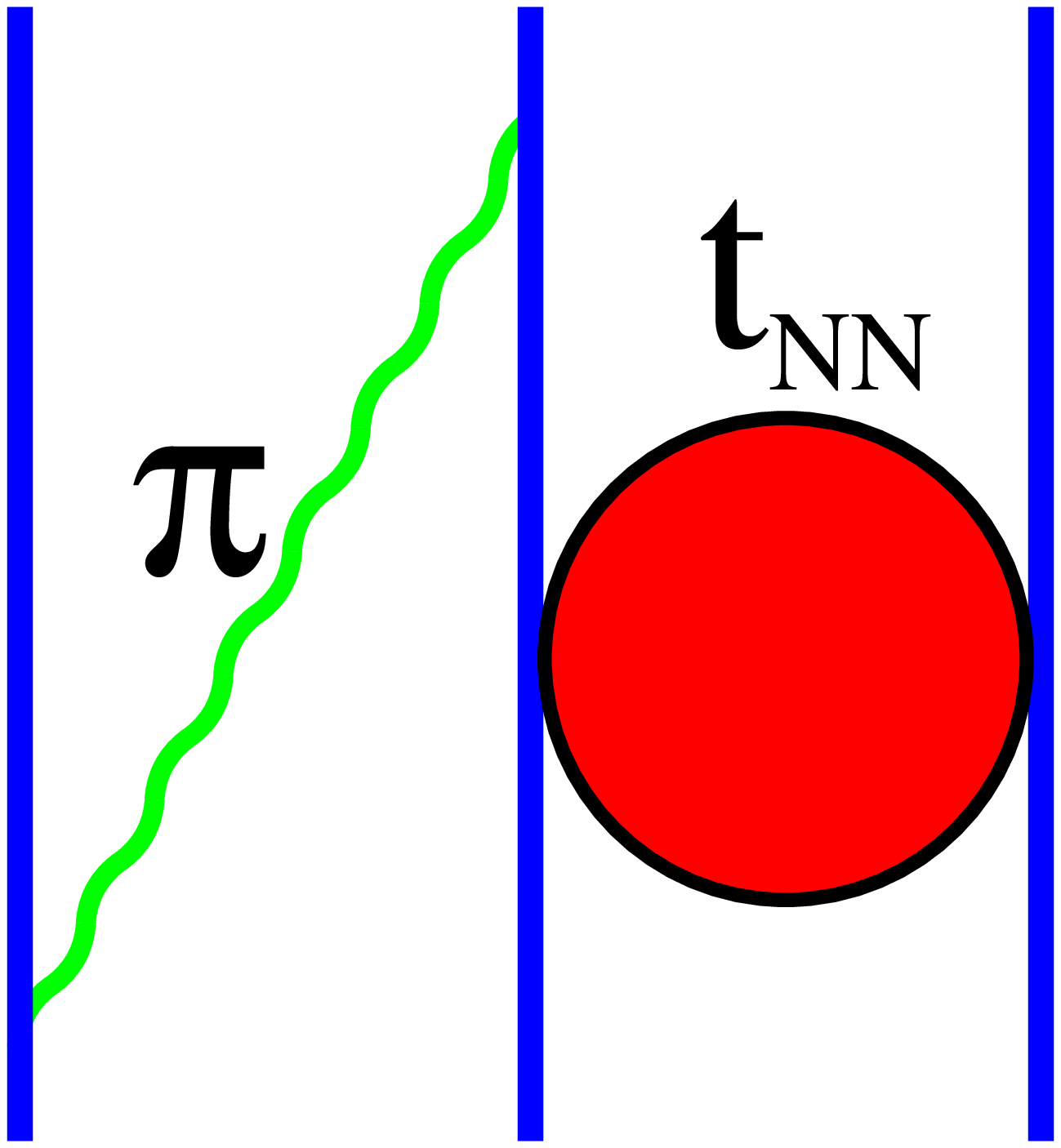}{}
}
\vspace{-0.7cm}
\caption[ ]{Another irreducible 3$NF$ diagram. The 2$N$-corre\-lation diagram.}
\label{fig2}
\end{figure}

The  structure of the second class of diagrams
obtained by cooling down the pion from the $\pi$-3$N$
approach has been overlooked
in all modern force calculations: it corresponds 
to the graph represented in 
Fig.~\ref{fig2}.  This describes
a complete correlation between
one of the two nucleons exchanging the meson
and the third one while the pion is ``in flight".
These diagrams should not be confused with those 
originated by pure mesonic retardation effects,
analyzed and evaluated many years ago in a sequence of papers,
in Refs.~\cite{Brueckner54,Pask67,Yang74}.

As has been shown later in Refs.~\cite{Yang86,Coon86},
those dia- 

\begin{figure}
\vspace{-0.4cm}
\centerline{\hspace{0.8cm}
\Spicture{1.9 in}{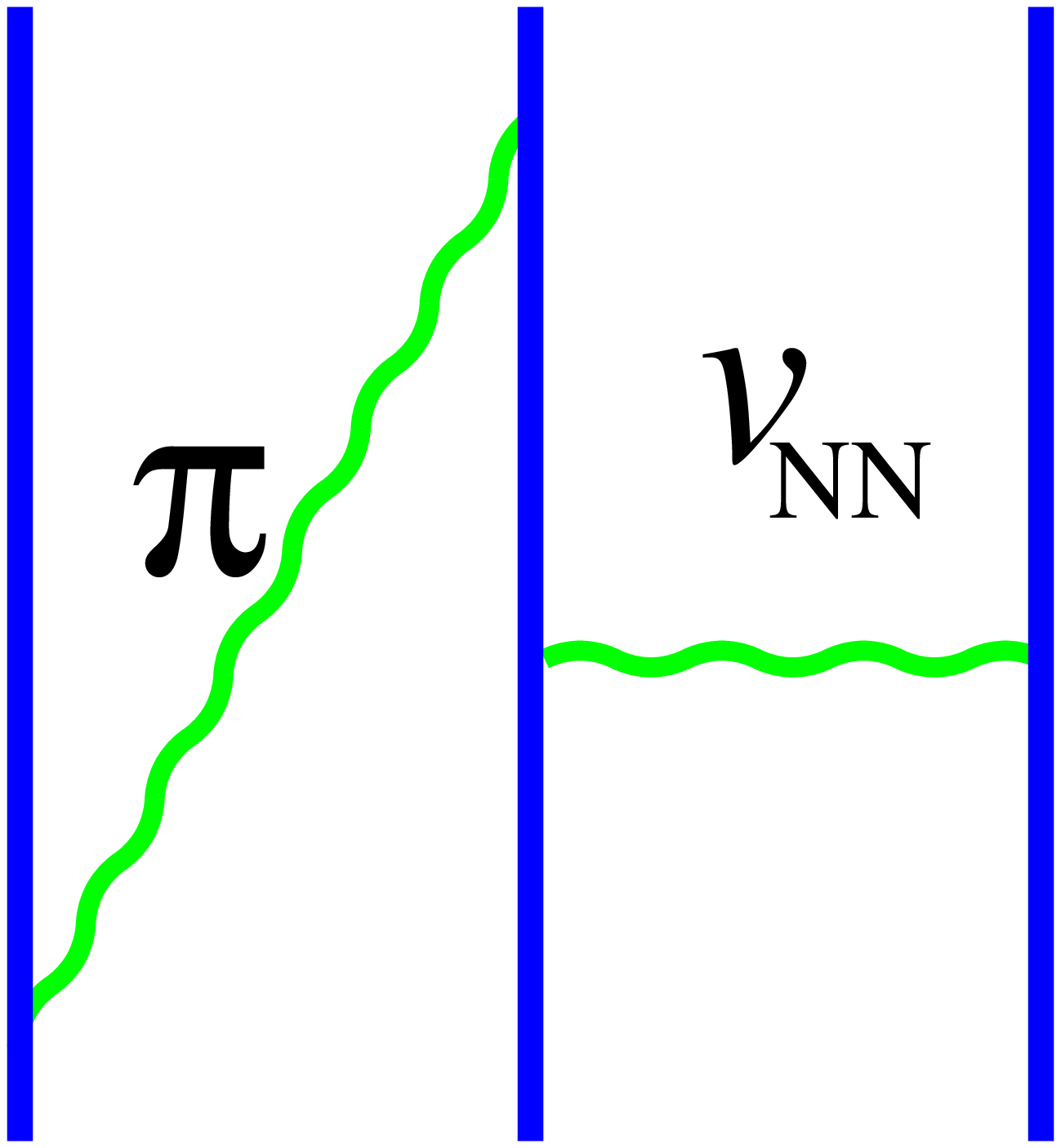}{}
\vspace{-0.7cm}
}
\caption[ ]{The irreducible Born diagram. This also must be subtracted}
\label{fig2born}
\end{figure}

\noindent
grams contributing to the 3$NF$ with pure mesonic retardation effects
must be excluded because they cancel out
against the corresponding contributions of 
pionic retardation effects arising from the Born term. 
This cancellation is correctly taken into account in the 
construction of modern 3$NF$ 
contributions~\cite{Friar99,Weinberg92,Epelbaoum98} and 
occurs also when considering 3$NF$ retardation effects coming from the 
exchange of heavier mesons~\cite{Eden96,Yang86}.
To take  into account the effects of this cancellation
we subtract from the diagram in Fig.~\ref{fig2}
the 2$^{nd}$ (irreducible) Born diagram shown in Fig.~\ref{fig2born}.

The reason for this subtraction can be understood in the following way:
The diagram in Fig.~\ref{fig2} considers
an irreducible contribution wherein the pion
propagates while a 2$N$ subsystem {\em clusterizes}. 
Obviously, such an effect does happen above the pion threshold,
since the reactions $N d\rightarrow N d \pi$ are observed;
the problem is to determine up to what extent this mechanism
is relevant at lower energies, where 
the $N+(NN)+\pi$ channel
is asymptotically closed but may still be important as an 
intermediate state.
The 3$NF$ mechanism of Fig.~\ref{fig2} is dynamically more complete
than the one shown in Fig.~\ref{fig2born} which has been demonstrated to
cancel out against the mesonic 
retardation corrections of the twice iterated Born term of 
Fig.~\ref{fig1born} (see Refs.~\cite{Yang86,Eden96} for details). 
Therefore, it is the difference between the two 
diagrams in Fig.~\ref{fig2} and \ref{fig2born} that survives from the
mesonic retardation effects and generates a new 3$NF$ contribution.
Had we replaced the full two-body $t$-matrix in this diagram by the
input potential (this corresponds to an ``instantaneous'', Born-type
approximation) then the cancellation would be matched exactly
and this 3$NF$ effect would disappear. Hence, \linebreak

\begin{figure}
\vspace{0.4cm}
\centerline{\hspace{1.3cm}
\Spicture{1.9 in}{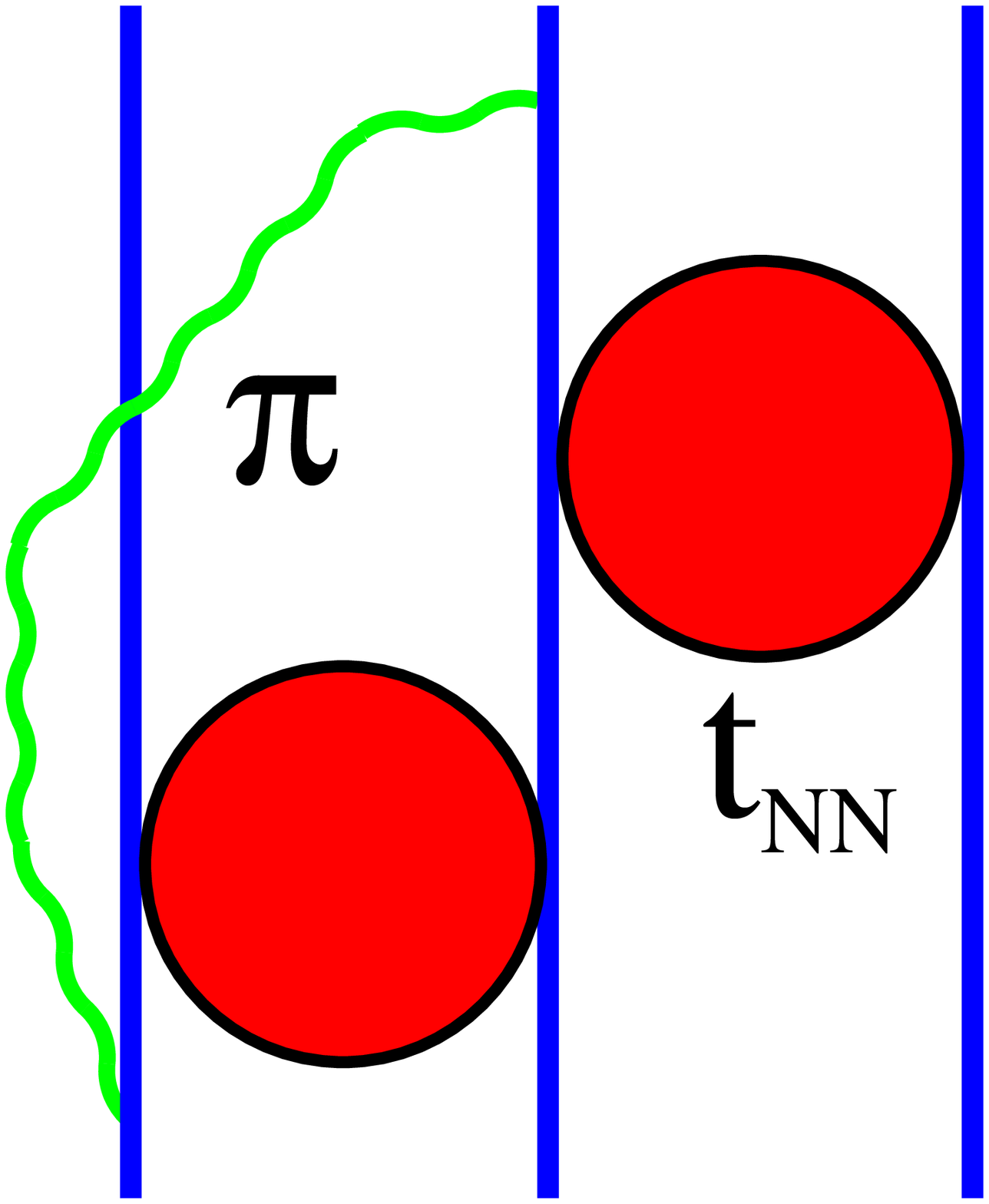}{}
}
\vspace{-0.5cm}
\caption[ ]{Yet another irreducible 3$NF$ diagram. The 
connected-correlation diagram.}
\label{fig3}
\end{figure}

\noindent
because of this incomplete cancellation, the effect 
{\em must} be entirely attributed to the energy dispersion 
of the intermediate 2$N$ correlation, 
combined with a ``long leg pion" diagram. 

Finally, it is possible to generate from the study of $Z_{ab}^{3N}$
other, more complicated, diagrams.  In Fig.~\ref{fig3} it is shown
just one example.  One class embraces all possible {\it connected}
3$N$ correlations while the exchanged pion is ``in flight".  On the
contrary, the diagrams of Fig.~\ref{fig2} represents all possible {\it
disconnected} 2$N$ correlations while the pion is ``in flight".

We summarize this section with some comments.
First, 
the analysis performed in this section was based on a new 
approach developed for the description of the pion 
dynamics within a nonrelativistic multinucleon context. 

The more practical approach in Ref.~\cite{Canton2000a} is sufficiently
systematic to generate three topologically different 
structures of irreducible diagrams for the 3$N$ force.
The more general approach of Ref.~\cite{Canton98}
will generate additional classes of irreducible (and reducible) 3$NF$
diagrams.  
The first diagrammatic structure that emerged is well known and 
is practically the only one explored
in modern few-nucleon calculations. 
Then there are irreducible diagrams whose topological structure
was not known. Most interesting is the class of diagrams
considering a 2$N$ correlation while the pion is being
exchanged. These diagrams should not be confused with the 
mesonic retardation effects, which produce a net 
cancellation amongst themselves. To our knowledge,
this contribution to the 3$NF$ has never been considered before
and will be analyzed in the next two sections of this paper.
Finally, we have also revealed the presence of another class based on 
connected 3$N$ correlations while the pion is being exchanged.
From the connected-kernel approach considered in Ref.~\cite{Canton98},
other more complicated classes could be produced. 
Their basic building blocks, however, are always the 2$N$ 
$t$-matrices and the $\pi N$ amplitudes, and therefore it is clear that
the two fundamental ingredients for the construction
of the 3$NF$ are the ones shown in  
Figs.~\ref{fig1} and \ref{fig2}.

\begin{figure}
\centerline{\hspace{1.cm}
\Spicture{3.2 in}{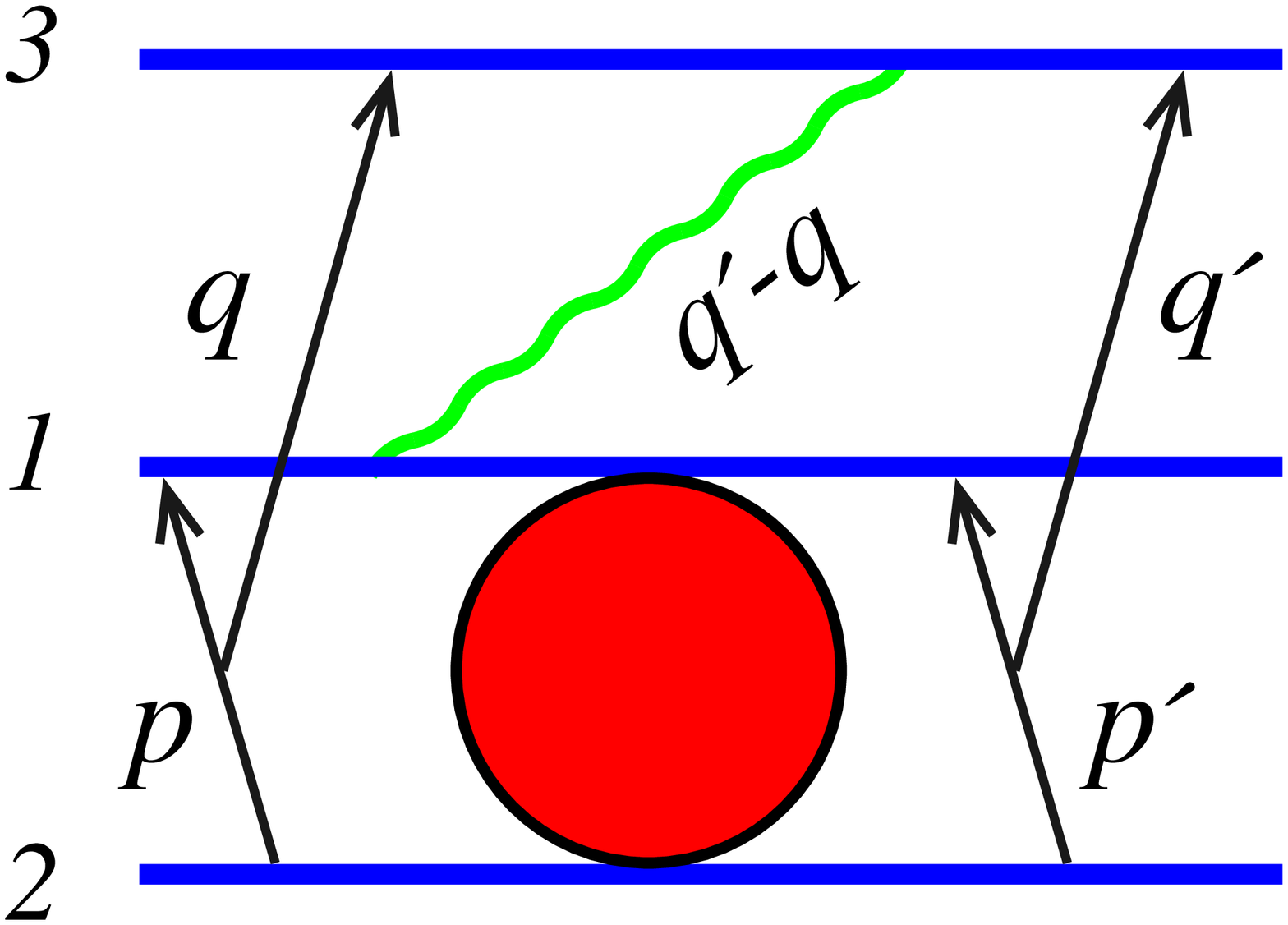}{}}
\vspace{-7pt}
\caption[ ]{Notation on the 3$N$ Jacobi momenta used
in the text.}
\label{Jacobi}
\end{figure}

\vspace{-9pt}

\section{ The ``odd" contribution to the 3$N$ force  }

In this section, we will first derive a 3$N$ interaction
from the irreducible 3$N$-force diagram shown in Fig.~\ref{fig2}.
Moreover, we will discuss how this irreducible 
3$NF$ produces a contribution acting in the triplet
odd-states for the 2$N$ subsystem.

To derive a 3$N$ interaction from the 
diagram in Fig.~\ref{fig2},
we sum all the possible diagrams corresponding to
a correlation between nucleons ``1" and ``2"
while the pion is ``in flight". There are four
of such diagrams and their sum provides the contribution
for an irreducible 3$N$ force 
in one single Faddeev component; namely the component
where nucleon ``3" represents the spectator. 
(This scheme of diagrammatic re-summation is a very natural
and automatic consequence of the formalism 
discussed in Ref.~\cite{Canton98}.)
We denote the resulting component of the 3$N$ force
as $V^{3N}_3$. The complete 3$N$ interaction obviously will result
from the sum over all three Faddeev components,
or equivalently from all diagrams resulting from 
the cyclic permutations of the nucleons in the four 
diagrams mentioned above.
Hence the total 3$NF$ contribution will result from
$V^{3N}=V^{3N}_1+V^{3N}_2+V^{3N}_3$.

The component $V^{3N}_3$ must be calculated 
according to the expression:
\begin{eqnarray}
V^{3N}_3 &=&
f_1G_0^{(4)}\tilde t_{12}G_0^{(4)}f_3^\dagger
+
f_2G_0^{(4)}\tilde t_{12}G_0^{(4)}f_3^\dagger \nonumber \\
& & +
f_3G_0^{(4)}\tilde t_{12}G_0^{(4)}f_1^\dagger
+
f_3G_0^{(4)}\tilde t_{12}G_0^{(4)}f_2^\dagger \, ,
\label{3NF-component}
\end{eqnarray}
where $f_1$ ($f_1^\dagger$) represents the $\pi NN$ vertex interaction
for exchanged pion production (absorption) on nucleon ``1",
$G_0^{(4)}$ denotes the intermediate propagation of the three nucleons plus
the exchanged pion, $\tilde t_{12}$ represents the subtracted 2$N$ 
$t$-matrix,
describing the correlation between nucleons ``1" and ``2" while the 
pion is ``in  flight". One must observe right from the start
that one cannot identify this amplitude with
the on-shell 2$N$ $t$-matrix. Indeed this subtracted amplitude enters off-shell
in the diagram, and with an energy shift. And below pion production
threshold, $\tilde t_{12}$ must be real because the free Green's function 
$G_0^{(4)}$ is not singular in this region of the real axis.

If one now considers the first of these four diagrams
with the Jacobi coordinates depicted as in Fig.~\ref{Jacobi},
using the static approximation 
and assuming the process in the c.o.m. 
(center of mass) of the system,
one obtains that this diagram corresponds to the following
contribution:
\begin{eqnarray}
D_1&=& \sum_\alpha 
{f_{\pi NN}\over m_\pi} 
{{\mbox{\boldmath $\sigma_1$}}\cdot ({\bf q-q'}) \over \sqrt{(2\pi)^3 2\omega}}
\tau_1^{\alpha} \,
G_0^{(4)}  \\
&& \!\!\!\!\!\!\!\!\times
{\tilde t}_{12}(p,p';E-{q^2\over 2\nu}-\omega_\pi) \,
G_0^{(4)}
{f_{\pi NN}\over m_\pi}
{{\mbox{\boldmath $\sigma_3$}}\cdot ({\bf q-q'}) \over \sqrt{(2\pi)^3 2\omega}}
\tau_3^{-\alpha}\, . \nonumber
\end{eqnarray}
With the sum over $\alpha$ it is intended that
all three isospin components of the pion field are summed up
(in pseudospherical representation), while
$\nu$ is the reduced mass of the spectator nucleon with respect
to the c.o.m. of the pair.

To derive the final expression of this diagram, we will  use the
static approximation. 
Originally, the Green's function on the left should read
\begin{equation}
G_0^{(4)}={1\over E- {p^2\over 2\mu} - {q^2\over 2\nu} -\omega_\pi} \, ,
\end{equation}
and similarly the one on the right should be
\begin{equation}
G_0^{(4)}={1\over E- {{p'}^2\over 2\mu} - {q^2\over 2\nu} -\omega_\pi} \, .
\end{equation}
Using the static approximation, we assume that
\begin{equation}
E\simeq {p^2\over 2\mu} + {q^2\over 2\nu}  \simeq 
 {{p'}^2\over 2\mu} + {q^2\over 2\nu}  \, .
\end{equation}
In this case,
both Green's functions on the left and right 
of $\tilde t_{12}$ can be approximated by the same
expression, namely
\begin{equation}
G_0^{(4)}\simeq -{1\over \omega_\pi} .
\end{equation}
We introduce also
${\bf Q}$ as the momentum transferred by the pion,
${\bf Q = q'-q}$, hence $\omega_\pi=\sqrt{m_\pi^2 + Q^2}$.

The subtracted $t$-matrix is estimated according to the expression
\begin{eqnarray}
\label{sottrazione}
\lefteqn{\tilde t_{12}(p,p';E-{q^2\over 2 \nu}-\omega_\pi(Q))} \nonumber \\
&& \; = t_{12}(p,p';E-{q^2\over 2 \nu}-\omega_\pi(Q)) -v_{12}(p,p')\, ,
\end{eqnarray}

\vfill
\noindent
where the potential-like term $v_{12}(p,p')$ contains
only OPE/OBE-type diagrams. 
The quantity $\tilde t_{12}$ depends on the Jacobi momenta
${\bf p}$, ${\bf p'}$, ${\bf q}$, and ${\bf q'}$ 
in a complicated way. The important feature, however, is that the 
energy of the 2$N$ subsystem is shifted to negative values
by the spectator kinetic energy {\em and} by the mesonic term
$\omega_\pi$.
In case of heavier mesons ($\omega_x \gg \omega_\pi$)
the 2$N$ energy becomes so negative (large in absolute value) 
that $t_{12}$ is close to $v_{12}$,
and hence the two quantities almost cancel each others.
As we will see further on,
with the pion the results are quite different because this 
meson - the Goldstone boson of the underlying chiral theory - is so light.

One has to repeat the same derivation also for the other three 
diagrams and sum over all the contributions. 
As a result, the third Faddeev component of the
irreducible 3$N$ force generated by the four terms
given by Eq.~(\ref{3NF-component}), can be expressed
as
\begin{eqnarray}
\label{3NF-v1}
\lefteqn{V^{3N}_3 =\ {f_{\pi NN}^2\over m_\pi^2}{1\over (2\pi)^3}}
 \\          
&& \! \times \!\left [
{
({\mbox{\boldmath $\sigma_1$}}\cdot{\bf Q})
({\mbox{\boldmath $\sigma_3$}}\cdot{\bf Q})
({\mbox{\boldmath $\tau_1$}}\cdot {\mbox{\boldmath $\tau_3$}})
+
({\mbox{\boldmath $\sigma_2$}}\cdot{\bf Q})
({\mbox{\boldmath $\sigma_3$}}\cdot{\bf Q})
({\mbox{\boldmath $\tau_2$}}\cdot {\mbox{\boldmath $\tau_3$}})
\over
\omega_\pi^2
}
\right ]  \nonumber \\
&&\times \, {
\tilde t_{12}(p,p';E-{{q}^2\over 2\nu} - \omega_\pi)
\over 
2\omega_\pi}
\nonumber \\
&& +{f_{\pi NN}^2\over m_\pi^2}{1\over (2\pi)^3}
\, {
\tilde t_{12}(p,p';E-{{q'}^2\over 2\nu} - \omega_\pi)
\over 
2\omega_\pi} 
\nonumber   \\ 
& & \!\!
 \times \! \left [
{ 
({\mbox{\boldmath $\sigma_1$}}\cdot{\bf Q})
({\mbox{\boldmath $\sigma_3$}}\cdot{\bf Q})
({\mbox{\boldmath $\tau_1$}}\cdot {\mbox{\boldmath $\tau_3$}})
+
({\mbox{\boldmath $\sigma_2$}}\cdot{\bf Q})
({\mbox{\boldmath $\sigma_3$}}\cdot{\bf Q})
({\mbox{\boldmath $\tau_2$}}\cdot {\mbox{\boldmath $\tau_3$}})
\over
\omega_\pi^2
}
\right ] \!, \nonumber
\end{eqnarray}
and the total contribution to the 3$N$ force 
will be given by summing up this contribution
together with those obtained from this 
by cyclic permutations of the three nucleons.

This formula implies several aspects on which 
we would like to comment.
 
{\em i)} We have used the nonrelativistic reduction of the $\pi NN$ vertex.
{\em ii)} We have neglected nucleon recoil effects on the basis
that the pion mass $m_\pi$ is much smaller than the nucleon mass, $M$.
{\em iii)} We have made use of the static approximation, implying 
that the pion 
``in flight" exchange momentum but not energy 
with the nucleon ``3".
{\em iv)} The form of the force obtained above is made symmetrical by 
the combined sum of the four diagrams, and below pion threshold the resulting expression
is Hermitian, being $\tilde t_{12}$ real.

It may be convenient to
transform $V_3^{3N}$ introducing the spin and isospin operators
for the 2$N$ subsystem, 
${\bf S}_{12}={\mbox{\boldmath $\sigma_1$}}+{\mbox{\boldmath $\sigma_2$}}$,
and 
${\bf T}_{12}={\mbox{\boldmath $\tau_1$}}+{\mbox{\boldmath $\tau_2$}}$,
respectively.
Then it is a matter of simple algebraic manipulations
to rewrite Eq.~(\ref{3NF-v1}) in the following form
\begin{eqnarray}
\label{3NF-v2}
V^{3N}_3=&& 
{ 
{f_{\pi NN}^2} 
\over 
{2 m_\pi^3} 
}
{
1
\over 
(2\pi)^3
}
{
({\bf S}_{\bf 12}\cdot{\bf Q})
({\mbox{\boldmath $\sigma_3$}}\cdot{\bf Q})
\over 
m_\pi^2+Q^2
}
({\bf T}_{\bf 12}\cdot {\mbox{\boldmath $\tau_3$}}) \nonumber \\
&& \;\; \times \,  \tilde t_{12}(p,p';E-{{q}^2\over 2\nu} - m_\pi)  
\nonumber \\
+& & 
{ 
{f_{\pi NN}^2} 
\over 
{2 m_\pi^3} 
}
{
1
\over 
(2\pi)^3
}
\, \tilde t_{12}(p,p';E-{{q'}^2\over {2\nu}} - m_\pi)  \nonumber \\
& & \;\; \times \, {
({\bf S}_{\bf 12}\cdot{\bf Q})
({\mbox{\boldmath $\sigma_3$}}\cdot{\bf Q})
\over 
m_\pi^2+Q^2
}
({\bf T}_{\bf 12}\cdot {\mbox{\boldmath $\tau_3$}})
\nonumber \\
-& &
{
{f_{\pi NN}^2}
\over 
2 m_\pi^3
}
{
1
\over 
(2\pi)^3
}
{
({\mbox{\boldmath $\sigma_1$}}\cdot{\bf Q})
({\mbox{\boldmath $\sigma_3$}}\cdot{\bf Q})
\over 
m_\pi^2+Q^2
}
({\mbox{\boldmath $\tau_2$}}\cdot {\mbox{\boldmath $\tau_3$}}) \nonumber \\
& & \;\; \times \,{
\, \tilde t_{12}(p,p';E-{{q}^2\over {2\nu}} - m_\pi) } 
\nonumber \\
- & &
{
{f_{\pi NN}^2}
\over 
2 m_\pi^3
}
{
1
\over 
(2\pi)^3
}
\, \tilde 
t_{12}(p,p';E-{{q'}^2\over {2\nu}} - m_\pi)  \nonumber \\
& & \;\; \times \, {
({\mbox{\boldmath $\sigma_1$}}\cdot{\bf Q})
({\mbox{\boldmath $\sigma_3$}}\cdot{\bf Q})
\over 
m_\pi^2+Q^2
}
({\mbox{\boldmath $\tau_2$}}\cdot {\mbox{\boldmath $\tau_3$}})
\nonumber \\
-& &
{
{f_{\pi NN}^2}
\over 2 m_\pi^3
}
{
1
\over 
(2\pi)^3
}
{
({\mbox{\boldmath $\sigma_2$}}\cdot{\bf Q})
({\mbox{\boldmath $\sigma_3$}}\cdot{\bf Q})
\over 
m_\pi^2+Q^2
}
({\mbox{\boldmath $\tau_1$}}\cdot {\mbox{\boldmath $\tau_3$}}) \nonumber \\
& & \;\; \times \, {
\, \tilde 
t_{12}(p,p';E-{{ q}^2\over {2\nu}} - m_\pi) } 
\nonumber \\
-& &
{
{f_{\pi NN}^2}
\over 2 m_\pi^3
}
{
1
\over 
(2\pi)^3
}
\, \tilde 
t_{12}(p,p';E-{{ q'}^2\over {2\nu}} - m_\pi)  \nonumber \\
& & \;\; \times \, {
({\mbox{\boldmath $\sigma_2$}}\cdot{\bf Q})
({\mbox{\boldmath $\sigma_3$}}\cdot{\bf Q})
\over 
m_\pi^2+Q^2
}
({\mbox{\boldmath $\tau_1$}}\cdot {\mbox{\boldmath $\tau_3$}})
\, . 
\end{eqnarray}
Here, we have also approximated
the two normalization factors of the pion field,
i.e., the two square roots in the denominators,
as
\begin{equation}
{1\over \sqrt{2\omega}\sqrt{2\omega} }\simeq {1\over 2 m_\pi} \, ,
\end{equation}
and we have shifted the 2$N$ amplitudes by the pion mass $m_\pi$,
in place of $\omega_\pi$. 
These approximations are really not essential, but they simplify
considerably the formulas of the partial 
wave expansions, without really altering the physics.
It is clear however that a more consistent calculation requires
the employment of the exact expressions.

Both expressions Eqs. (\ref{3NF-v1})-(\ref{3NF-v2}) 
are symmetrical at sight. The latter one has the advantage
that it is possible to isolate from the rest the contribution
given by the first two terms,
which we rewrite as
\begin{eqnarray}
\label{3NF*}
V^*_3&=&
{ 
{f_{\pi NN}^2} 
\over 
{2 m_\pi^3} 
}
{
1
\over 
(2\pi)^3
}
{
({\bf S}_{\bf 12}\cdot{\bf Q})
({\mbox{\boldmath $\sigma_3$}}\cdot{\bf Q})
\over 
m_\pi^2+Q^2
}
({\bf T}_{\bf 12}\cdot {\mbox{\boldmath $\tau_3$}}) \nonumber \\
& & \times \, 
\tilde t_{12}(p,p';E-{{q}^2\over 2\nu} - m_\pi)  + h.c. 
\end{eqnarray}
The remaining part of $V^{3N}_3$,  i.e., the sum over
the last four terms, mixes together the spin components of nucleon
``1" with the isospin components of nucleon ``2".
We will not discuss further this contribution
and leave it for future studies. The last four terms in Eq. (\ref{3NF-v2})
is just one of the many irreducible contributions which should
be added up to build the full spin-isospin structure of the 3$N$ interaction.

In the following, we will focus the attention
on $V^*_3$, which has an interesting spin-isospin structure
since it depends only on the spectator and pair coordinates.
Because of the presence of the spin-isospin operators
${\bf S_{12}}$ and ${\bf T_{12}}$,
such term vanishes unless the 2$N$ pair is in a triplet state
for {\em both} spin and isospin coordinates. And since
the nucleon pair must be in an antisymmetric state
because of the generalized Pauli principle, then the allowed 
orbital momentum of the pair can be only odd. 
This means that this contribution to the irreducible 3$N$ force
acts only in triplet odd states ($^3P_J$-waves, $^3F_J$-waves, etc.) 
of the two-nucleon subsystem. Stated in other words,
the third nucleon, by means of this contribution modifies
selectively the triplet odd states of the 2$N$ subsystem with respect
to a free, isolated nucleon-nucleon pair. We believe that
this mechanism can possibly modify those observables
particularly sensitive to the triplet $p$- and $f$-waves and might 
therefore affect also the nucleon-deuteron vector analyzing powers.

\section{ Partial wave decomposition }

We provide the partial-wave decomposition of $V^*_3$.
We will work in the so-called channel spin 
coupling since this is the most natural scheme
for nucleon-deuteron scattering.

The channel-spin coupling is defined according
to the following notation
\begin{equation}
w = (((ls)j\sigma)KL)\Gamma\Gamma_z ,
\end{equation}
where $l$, $s$, and $j$ represent
respectively the orbital momentum, spin, and
total spin-angular momentum of the pair (hence,
$s$ represents the quantum number associated to
the operator ${\bf S_{12}}$). The total spin of the pair 
$j$ is then coupled with the intrinsic spin of the spectator
$\sigma=1/2$ to provide the so-called channel spin $K$.
And finally, this is coupled to the orbital 
angular momentum of the third nucleon, $L$,
to give the total angular momentum of the 3$N$
system $\Gamma$ and its azimuthal component $\Gamma_z$.

We observe that $V^*_3$ has an interesting structure
in the pair-spectator coordinate system. The structure
is that of a OPE contribution, but in the spectator
coordinates, multiplied by a full 2$N$ interaction depending on 
the internal coordinates of the pair.
Given this structure, it is not so difficult to perform
the partial wave decomposition of  $V^*_3$.
Indeed, one can consider the spectator coordinates 
and  separate the 3$N$ potential into
a spin-spin component and a tensor one, following the 
standard procedure for the OPE term (see, e.g.,
Refs.~\cite{Ericson88,Machleidt87}),
\begin{eqnarray}
\lefteqn{{
({\bf S}_{\bf 12}\cdot{\bf Q})
({\mbox{\boldmath $\sigma_3$}}\cdot{\bf Q})
\over 
m_\pi^2+Q^2
}} \\
&&={1\over 3} 
\left [ -{m_\pi^2\over {m_\pi^2 + Q^2}}
({\bf S}_{\bf 12}\cdot {\mbox{\boldmath $\sigma_3$}})
+{ Q^2\over m_\pi^2+Q^2}
{\mbox{\boldmath $\Sigma_{12}$}}(\hat {\bf Q})
\right ] \, , \nonumber 
\end{eqnarray}
where $(\hat {\bf Q})$ is the angular part of 
the spectator momentum, and 
${\mbox{\boldmath $\Sigma_{12}$}}(\hat {\bf Q})=
3
({\bf S}_{\bf 12}\cdot\hat{\bf Q})
({\mbox{\boldmath $\sigma_3$}}\cdot\hat{\bf Q})
-({\bf S}_{\bf 12}\cdot {\mbox{\boldmath $\sigma_3$}})$ 
is the tensor operator.

In the above  equation, we have neglected the contact term,
on the ground that its contribution will be 
unavoidably smeared out
when taking into account the extended structure
of the sources of the meson field. The
extended nature of the sources 
have to be included in $V^*_3$ by means
of phenomenological $\pi NN$ form factors.
In a fully consistent calculation
these formfactors should be the same as those used
in the standard 2$N$ OPE contribution. Then
$V^*_3$ is completely fixed
by the full expression of the 2$N$ potential.

For the spin-spin part of $V^*_3$,
we obtain the following partial-wave decomposition:
\begin{eqnarray}
\lefteqn{\langle p,q,w| V^*_3(spin-spin)|p',q',w'\rangle
=
\delta_{ss'}
\delta_{s1}
\delta_{LL'}
\delta_{KK'}} \phantom{000}\nonumber \\
&& \times \,
\delta_{\Gamma\Gamma'}
\delta_{\Gamma^{}_z\Gamma'_z} \,
\hat j \hat {j'} \,
12 \,
{2\over \pi} \, 
I_{L}(q,q')  \nonumber \\
&&\times
(-)^{l+j+K+j'+{1\over2}}
\left\{ \matrix{1 & j & j' \cr
        l & 1 & 1} \right\} 
\left \{ \matrix{1 & j & j' \cr
        K & {1\over 2}  & {1\over 2} } \right\} 
 \nonumber \\
&& \times
\left [\tilde t_{12}(p,p';E-{q^2\over 2\nu}-m_\pi)\right]^{j's'}_{ll'}
{1\over 2m_\pi}    + h.c. 
\end{eqnarray}

\vfill

For the tensor-spin component the decomposition
in partial waves is slightly more complicated, being

\vfill

\begin{eqnarray}
\lefteqn{\langle p,q,w| V^*_3(tensor)|p',q',w'\rangle
=
\delta_{ss'}
\delta_{s1}
\delta_{\Gamma\Gamma'}
\delta_{\Gamma^{}_z\Gamma'_z}
\hat j \hat {j'}
\hat K \hat {K'} } \nonumber \\
&& \times \, \hat L \hat {L'}
12{2\over \pi}\,\sqrt{30} \,I_{LL'}(q,q')
\left(
\matrix{ L & 2 & L'\cr
         0 & 0 & 0}
\right)
(-)^{l+j+\Gamma+K'}
 \nonumber \\
&& \times 
\left \{ \matrix{1 & j & j' \cr
        l  & 1 & 1} \right\} 
\left \{ \matrix{K & 2 & K' \cr
        L'  & \Gamma   & L} \right\} 
\left \{ \matrix{1 & j & j' \cr
       1 & {1\over 2}&  {1\over 2} \vspace{1mm} \cr
       2 & K & K' }\right\} 
 \nonumber  \\
&&\times
\left [\tilde t_{12}(p,p';E-{q^2\over 2\nu}-m_\pi)\right]^{j's'}_{ll'}
{1\over 2m_\pi} + h.c. 
\end{eqnarray}

\vfill

The two quantities $I_{L}(q,q')$ and 
$I_{L,L'}(q,q')$ represent
well known 
Fourier-Bessel transforms
of Yukawa-type functions,
\begin{eqnarray}
\lefteqn{I_{L}(q,q')} \\
&& \;\;=- {f_{\pi NN}^2\over 12\pi}
\int_0^\infty j_L(qR) \, j_L(q'R)
\left( {e^{-m_\pi R}\over R}\right) R^2dR \, , \nonumber
\end{eqnarray}
and
\vspace{-1mm}
\begin{eqnarray}
I_{LL'}(q,q')
&=&- {f_{\pi NN}^2\over 12\pi}
\int_0^\infty j_L(qR) \,j_{L'}(q'R) \\
& \times& \left[ {e^{-m_\pi R}\over R} \left (1+{3\over m_\pi R}+
{3 \over (m_\pi R)^2} \right) \right] R^2dR \, . \nonumber 
\end{eqnarray}

\vfill

The resulting analytical expressions for these integrals
are well known~\cite{redish87}.

Finally, the potential matrix elements for the spin and tensor
parts must be multiplied by the isotopic component,
which is the same in both cases: \pagebreak

$ $ \vspace{-18mm}

\begin{figure}
\centerline{
\Spicture{4. in}{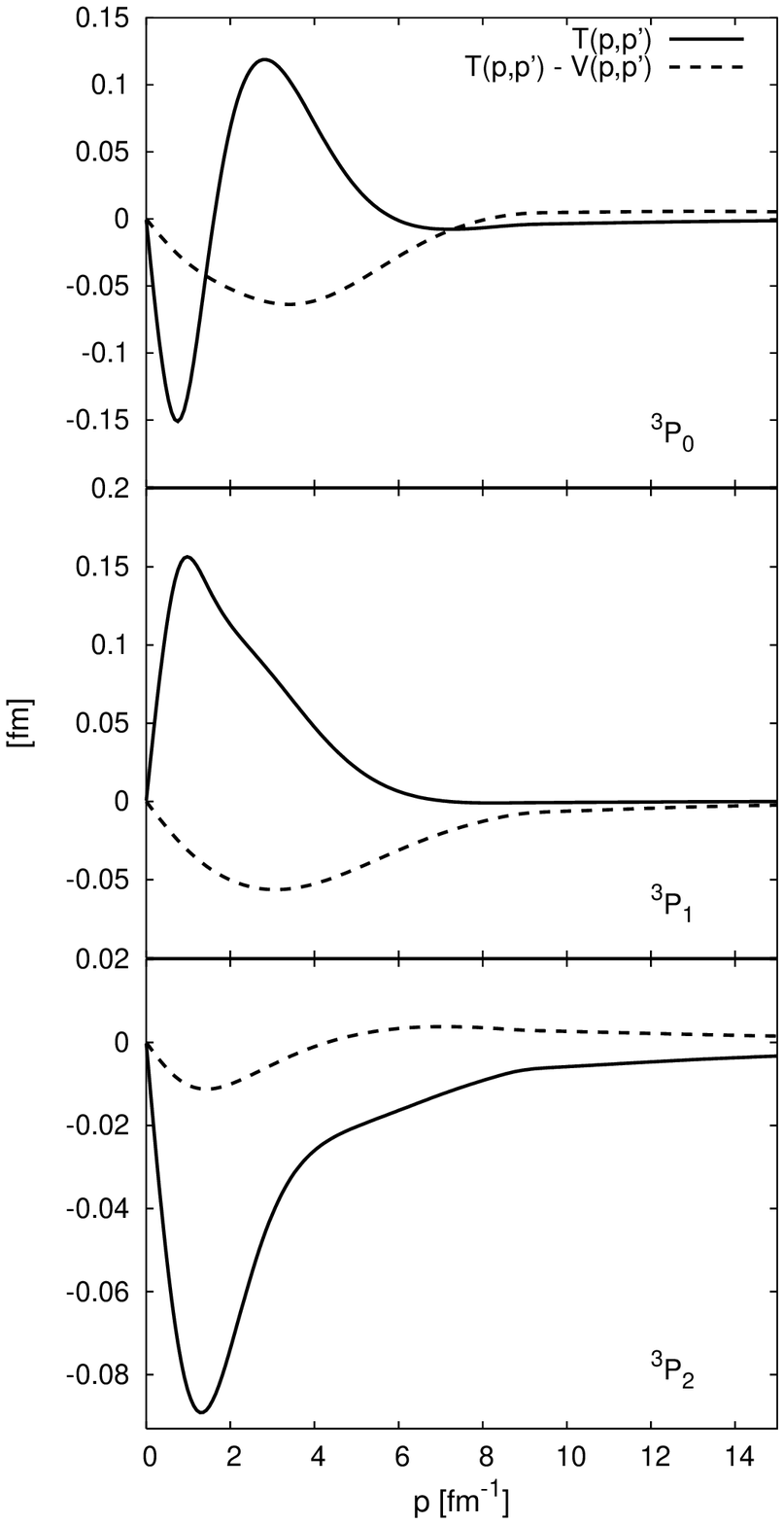}{}}
\vspace{-1.2cm}
\caption[ ]
{Effect of the subtraction in the triplet $p$-waves.
The full line represents the unsubtracted 2$N$ $t$-matrix
for E = -150 MeV and $p' = 0.89$ fm$^{-1}$ while the pion is
 ``in flight". The dashed line shows
the same amplitude when mesonic retardations 
(discussed in the text) are subtracted.
A 15-30\% effect survives from the cancellation.
This generates the new 3$NF$ component. }
\label{Diagrams}
\end{figure}

\vspace{2mm}

\begin{eqnarray}
\langle (t\tau)TT_z|
({\bf T}_{\bf 12} &\cdot& {\mbox{\boldmath $\tau_3$}})
|(t'\tau)T'T'_z\rangle=
\delta_{tt'} \,
\delta_{t1} \,
\delta_{TT'} \,
\delta_{T^{}_zT_z'}  \nonumber \\
&& \times 12 \, (-)^{{3\over 2}+T}
\left\{ 
\matrix{T & {1\over 2} & 1\cr
        1 & 1 & {1\over 2} }\right\} \, .
\end{eqnarray}
Here $\tau=1/2$ is the isospin of the spectator 
nucleon,
$t$ is the isotopic spin of the 2$N$ pair,
and $T$ is the total isospin of the 3$N$ system.   
As ought to be expected, 
only for isovector pairs this matrix element is nonzero.

Equation~(\ref{3NF*}) 
represents an OPE potential in the spec\-tator-pair coordinates, 
times a subtracted 2$N$ $t$-matrix for the internal coordinates
of the pair. This 2$N$ amplitude is quite off-shell,
because of the presence of a pion-exchange term, in addition to
the standard shift due to the spectator nucleon.
In the limit of a heavy meson exchange ($m_x\rightarrow 1$ GeV)
the energy of the 2$N$ subsystem will be large
and negative, and $t_{12}$ will be dominated by its 
Born OBE term $v_{12}$. Hence the effect of this 3$NF$ is suppressed
because approximately $\tilde t_{12}\simeq 0$. On the contrary,
as shown in Fig.~\ref{Diagrams}, for the lightest meson,
a 15-30 \% effect (at least) survives from the cancellation
and this generates the ``odd" contribution to the 3$NF$. 
The figure shows the comparison between 
the unsubtracted (solid line) and the subtracted (dashed line)
$t$-matrices, for a 2$N$ energy of -150 MeV, consistent with the 
calculation of the diagram shown in  Fig.~\ref{fig2}. 
The $t$-matrices have been calculated in the relevant triplet $p$-states, 
with the Bonn {\sl B} potential \cite{Machleidt87}, which is of OBE type.
Then the subtracted amplitude 
is given simply by Eq.~(\ref{sottrazione}).
The lines show the $t_{12}$ and $\tilde t_{12}$ amplitudes
as a function of the momentum $p$, while $p'$
was fixed at the value $p'=0.89$ fm$^{-1}$.
With other values of $p'$ we found the same effect.
Also, in order to show that the cancellation
could not produce an overall vanishing result
we made a more stringent test, by assuming
that when the mass of the particle ``in flight" is of the order of 
$\Lambda\simeq 1$ GeV, then the diagram of Fig.~\ref{fig2}
is canceled {\em exactly} against the mesonic retardation 
corrections of Fig.~\ref{fig2born}. In this case,
the subtracted $t$-matrix entering in the {\em pionic} diagram
can be evaluated according to the expression
\begin{eqnarray}
\label{sottrazione1}
\lefteqn{\tilde t_{12}(p,p';E-{q^2\over 2 \nu}-\omega_\pi(Q))} \nonumber \\
&& \;\;\approx t_{12}(p,p';E-{q^2\over 2 \nu}-\omega_\pi(Q)) -
t_{12}(p,p';-\Lambda) \, ,
\end{eqnarray}
and this expression can be employed with 
all types of phenomenological $NN$ potentials.
We checked $\tilde t_{12}$ for an energy of the 
2$N$ subsystem around -150 MeV, for the Bonn {\sl B} 
and for the Paris potential \cite{Lacombe80},
and with both interactions we found that a 
10-20 \% effect was surviving after the subtraction, thus providing 
evidence that the 
cancellation cannot hold exactly and simultaneously
in both cases of light- and heavy-meson exchanges.

However, it is also possible
to modify the behavior of the 2$N$ amplitude in this energy region, 
so that to obtain a 3$NF$ contribution of larger or smaller size, without 
obviously altering the constraints to the 2$N$ amplitudes 
from comparison with phase-shift analyses.

\section{ Summary and Conclusions}

\vfill

As discussed in the introduction,
discrepancies between theoretical calculations and experimental 
measurements give a clear indication that 3$NF$'s of new structure
are badly needed. The existing 3$NF$ models do not
seem to provide in full the correct spin-isospin
structure of the 3$N$ force.

In this paper, we have suggested a new mechanism which 
generates an irreducible 3$NF$ whose structure 
is topologically
very different from those explored up to now. Therefore,
the force generated by this mechanism should be considered 
as an additional contribution to the full 3$N$ interaction.
This mechanism is generated under the hypothesis that the 
standard few-nucleon dynamics and the pion-exchange processes 
are intertwined more strongly than what has been generally assumed
up to now. In particular, it is the intermediate formation
of a virtual 2$N$ cluster during a pion-exchange process
that gives rise to this new 3$NF$ term. In an instantaneous
approach, such a mechanism is 100\% suppressed because
of the presence of a well-known cancellation effect
which involves the meson-retardation corrections
of the reducible Born term of Fig.~\ref{fig1born} 
as well as all possible irreducible 3$NF$ diagrams obtained
by subsuming the exchange of two pions in their variety
of possible time orderings. It was known also that
the same cancellation occurs when considering similar
processes involving heavy-meson exchanges.

We have sized the effect of this cancellation
more precisely by considering a complete 2$N$ rescattering process,
not just its Born term (of Fig.~\ref{fig2born}).
This was possible by extrapolating 
the $NN$ $t$-matrix downwards to negative energy by a shift 
given by the spectator kinetic energy and by the relativistic energy 
of the meson. The result of this study reveals that a 15-30 \%
contribution survives from this cancellation in case
of a cluster formation while a pion is ``in flight". Instead,
the cancellation is more pronounced when considering the 
same process while a heavy meson in ``in flight".  
The reason for this difference
can be entirely attributed to the fact that the mass of the pion
is approximately comparable to the average momenta exchanged 
between nucleons in nuclear process, while the mass of the heavier 
mesons are larger.

We have studied the spin-isospin structure of the 3$NF$ generated by
this new, pion-induced mechanism and we have extracted an
important contribution which selectively operates in the
triplet-odd waves for the 2$N$ subsystem. 
While we acknowledge that the way to fully understand
the spin-isospin structure of the 3$N$ interaction 
is still long and difficult,
we conclude that this new term may
possibly contribute to this structure, especially
by affecting those spin observables most
sensible to the $^3P_J$ waves, such as the $Nd$ vector 
analyzing powers.

\acknowledgements

The work of L.~C. is performed under the Murst-PRIN Project 
``Fisica Teorica del Nucleo e dei Sistemi a Pi\'u Corpi".
The work of W.~Sch. is supported
by the Natural Science and Engineering Research Council of Canada.


\begin{references}

\vspace{-1.3cm}

\bibitem{Gloeckle96} 
W. Gl\"ockle, H. Wita{\l}a, D. H\"uber, H. Kamada, and J. Golak, 
Phys. Rep. {\bf 274}, 107 (1996).

\bibitem{pan3N}  {\it 
Proceedings of the XVth International Conference
on Few-Body Problems in Physics} (Edited by
J.~C.~S. Bacelar, A.~E.~L. Dieperink, R.~A. Malfliet, and L.~P. Kok,
Groningen, 1997); published in: Nucl. Phys. {\bf A631}, (1998).  

\bibitem{Polyzou90} W.~N. Polyzou and W. Gl\"ockle,
Few-Body Syst. {\bf 9}, 97, (1990).

\bibitem{Huber98} D. H\"uber, H. Wita{\l}a, H. Kamada, A. Nogga,
and W. Gl\"ockle, Nucl. Phys. {\bf A631}, 663c (1998).

\bibitem{Witala98} H. Wita{\l}a, W. Gl\"ockle, D. H\"uber,  J. Golak, 
and H. Kamada, Phys. Rev. Lett. {\bf 81}, 4820 (1998).

\bibitem{Nemoto98} S. Nemoto, K. Chmielewski, S. Oryu, and P.~U. Sauer,
Phys. Rev. C {\bf 58}, 2599 (1998).

\bibitem{Canton97} L. Canton and W. Schadow, Phys. Rev. C {\bf 56},
1231 (1997).

\bibitem{Canton98b} L.~Canton, G.~Cattapan, G.~Pisent,  W.~Schadow, 
and J.P.~Svenne, Phys. Rev. C {\bf 57}, 1588 (1998).

\bibitem{Canton99} L. Canton and W. Schadow, Phys. Rev. C {\bf 61},
064009 (2000) .

\bibitem{Kievsky96} A. Kievsky, S. Rosati, W. Tornow, and M. Viviani,
Nucl. Phys. {\bf A607}, 402 (1996).

\bibitem{Tornow98} W. Tornow and H. Wita{\l}a, Nucl. Phys. {\bf A637},
280 (1998).

\bibitem{Friar98} D. H\"uber and J. L. Friar, Phys. Rev. C 
{\bf 58}, 674 (1998).

\bibitem{NI93} V.~G.~J. Stoks, R.~A.~M. Klomp, M.~C.~M. Rentmeester, and 
J.~J. de Swart, Phys. Rev. C {\bf 48}, 792 (1993).

\bibitem{Canton98} L. Canton, Phys. Rev. C {\bf 58}, 3121  (1998).

\bibitem{AGS} E.~O. Alt, P. Grassberger, and W. Sandhas,
Nucl. Phys. {\bf B2}, 167 (1967).

\bibitem{Canton2000a} L. Canton, T. Melde, and J.~P. Svenne, nucl-th/0004045.

\bibitem{Yang86} S.~Y. Yang and W. Gl\"ockle, Phys. Rev. C {\bf 33}, 1774 
(1986).

\bibitem{Coon86} S.~A. Coon and J. L. Friar, Phys. Rev. C 
 {\bf 34}, 1060 (1986).

\bibitem{Weinberg92} S. Weinberg, Phys. Lett. B {\bf 295}, 114 (1992).

\bibitem{Sauer92} P.~U. Sauer, Nucl. Phys. {\bf A543}, 291c (1992).

\bibitem{F} L.~D. Faddeev, {\em Mathematical Aspects of the Three-Body
Problem in the Quantum Scattering Theory},
Israel Program for Scientific Translations, Jerusalem, 1965.

\bibitem{Y} O.~A. Yakubovsk{\u\i}, Sov. J. Nucl. Phys. {\bf 5}, 937 (1967).

\bibitem{GS} P. Grassberger and W. Sandhas,
Nucl. Phys. {\bf B2}, 181 (1967).

\bibitem{EST} D.~J. Ernst, C.~M. Shakin and R.~M.~Thaler, Phys. Rev. C 
{\bf 8}, 46 (1973).

\bibitem{Cornelius90} T. Cornelius, W. Gl\"ockle, J. Haidenbauer,
Y. Koike, W. Plessas, and H.~Wita{\l}a, Phys. Rev. C {\bf 41}, 2538
(1990). 

\bibitem{Parke91} W.~C.~Parke, Y.~Koike, D.~R.~Lehman, and L.~C.~Maximon,
Few-Body Syst. {\bf 11}, 89 (1991).

\bibitem{Schadow98} W.~Schadow, W.~Sandhas, J.~Haidenbauer, and A. Nogga, 
Few-Body Syst. {\bf 28}, 241 (2000).

\bibitem{Huber99} D. H\"uber, J.~L. Friar, A. Nogga, H. Wita{\l}a,
U. van Kolck, {\em nucl-th/9910034}. 

\bibitem{Friar99} J.~L. Friar, D. H\"uber, and U. van Kolck,
Phys. Rev. C {\bf 59}, 53 (1999).

\bibitem{FUMI} J.-I. Fujita and H. Miyazawa, Prog. Theor. Phys. {\bf 17},
360 (1957).

\bibitem{TM} S.~A. Coon, M.~D. Scadron, P.~C. McNamee, B.~R. Barrett,
D.~W.~E. Blatt, and B.~H.~J. McKellar, Nucl. Phys. {\bf A317}, 242 (1979).

\bibitem{BR} H.~T. Coelho, T.~K. Das, and M.~R. Robilotta, Phys. Rev. C 
{\bf 28}, 
1812 (1983); M.~R. Robilotta and H.~T. Coelho, Nucl. Phys. {\bf A460}, 
645 (1986).

\bibitem{Eden96} J.~A. Eden and M.~F. Gari, Phys. Rev. C {\bf 53},
1510 (1996).

\bibitem{TX} U. van Kolck, Phys. Rev. C {\bf 49}, 2932 (1994).

\bibitem{UA} J. Carlson, V.~R. Pandharipande, and R.~B. Wiringa,
Nucl. Phys. {\bf A401}, 59 (1983); B.~S. Pudliner, V.~R. Pandharipande, 
J. Carlson, and R.~B. Wiringa, Phys. Rev. Lett. {\bf 74}, 4396 (1995).

\bibitem{Brueckner54} K. A. Brueckner, C. Levinson, and H. Mahmoud, Phys. Rev. 
{\bf 95}, 217 (1954).

\bibitem{Pask67} C. Pask, Phys. Lett. {\bf 25B}, 78 (1967).

\bibitem{Yang74} S.~Y. Yang, Phys. Rev. C {\bf 19}, 2067 (1974).

\bibitem{Epelbaoum98} E. Epelbaoum, W. Gl\"ockle, and Ulf-G. Mei{\ss}ner,
Nucl. Phys. {\bf A637}, 107 (1998).

\bibitem{Ericson88} T. Ericson and W. Weise, {\em Pions and Nuclei},
Clarendon Press, Oxford, 1988.

\bibitem{Machleidt87} R. Machleidt, K. Holinde, and Ch. Elster,
Phys. Rep. {\bf 149}, 1 (1987).

\bibitem{redish87} E.~F. Redish and K. Stricker-Bauer, 
Phys. Rev. C {\bf 36}, 513 (1987).

\bibitem{Lacombe80} M.~Lacombe, B.~Loiseau, J.~M.~Richard and
       R.~Vinh Mau, Phys. Rev. C {\bf 21}, 861 (1980).

\end{references}
\end{document}